\documentclass[prl,reprint,amsmath,amssymb,superscriptaddress]{revtex4-1}

\usepackage{color}

\usepackage{graphicx}

\usepackage[english]{babel}

\usepackage{mathrsfs}

\usepackage[normalem]{ulem}

\newcommand{\vct}[1]{\mathbf{#1}}
\newcommand{\kb}[0]{k_{\textup{B}}}

\newcommand{\ci}[0]{c_i^{l_0}}

\begin{document}

\title{Microscopic Picture of Cooperative Processes in Restructuring Gel Networks}
\author{Jader Colombo}
\affiliation{Department of Civil, Environmental and Geomatic Engineering, ETH Z\"{u}rich, CH-8093 Z\"{u}rich, Switzerland}
\author{Asaph Widmer-Cooper}
\affiliation{School of Chemistry, University of Sydney, NSW 2006, Australia}
\author{Emanuela Del Gado}
\affiliation{Department of Civil, Environmental and Geomatic Engineering, ETH Z\"{u}rich, CH-8093 Z\"{u}rich, Switzerland}

%\date{\today}

\begin{abstract}
Colloidal gel networks are disordered elastic solids that can form even in extremely dilute particle
suspensions. With interaction strengths comparable to the thermal energy, their stress-bearing
network can locally {\it restructure} via breaking and reforming interparticle bonds. This allows
for yielding, self-healing and adaptive mechanics under deformation. Designing such features
requires controlling stress-transmission through the complex structure of the gel and this is
challenging because the link between local restructuring and overall response of the network is
still missing.
 %bears enormous potential for designing responsive nano-composites, 
%but it is challenging because stress-transmission in such complex structures is far from trivial.
%In particular, it is still not understood how single bond changes affect the gel network and its dynamics.
Here, we use a space resolved analysis of dynamical processes and numerical simulations of a model
gel to gain insight into this link. We show that consequences of local bond-breaking propagate along
the gel network over distances larger than the average mesh size.  This provides the missing
microscopic explanation for why non-local constitutive relations are necessary to rationalize the
non-trivial mechanical response of colloidal gels.
\end{abstract}

%\pacs{}

\maketitle

%% {\bf Colloidal gel networks are disordered elastic solids that can form even in extremely dilute particle suspensions  
%% %Colloidal suspensions of nano-particles solidify even if very diluted, because short-range interactions make particles aggregate into thin stress-bearing structures
%% \cite{trappe_nature2001, plu_nature}. 
%% With interaction strengths of a few $k_{B}T$, the stress-bearing network can locally {\it restructure} via breaking and reforming interparticle bonds. This allows for yielding, self-healing and adaptive mechanics under deformation. Controlling the link between local restructuring and mechanical response bears enormous potential for designing responsive nano-composites, but it is challenging because stress-transmission in such complex structures is far from trivial.
%% In particular, it is still not understood how single bond changes affect the gel network and its dynamics. Here, using numerical simulations of a model gel and a space resolved analysis of 
%% the relationship between its structure and dynamics, we show how local bond-breaking induces cooperative relaxation of the gel network over distances up to a few times the average mesh size.
%% This provides explicit microscopic insight into why non-local constitutive relations are necessary to rationalize the non-trivial mechanical response of colloidal gels.}

%\section{Introduction}
Colloidal suspensions can solidify even if very diluted, because short-range interactions make
particles aggregate into thin stress-bearing structures \cite{trappe_nature2001, plu_nature}.  As
for other network-forming soft materials, including many with important biological functions or
technological potential \cite{lieleg_nmat2011,pochan_softmatter2010}, there is still very limited
understanding of how local restructuring processes ultimately affect the transmission of stress
through the complex microstructure of the network, where weakly connected or soft regions may
coexist with stiffer domains. Experiments suggest that cooperative dynamical processes are
responsible for the complex mechanics of gels, which combines both liquid-like and solid-like
features and can only be rationalized by non-local constitutive relations
\cite{luca_prl2009,maccarrone,bocquet_nature}, however a microscopic explanation for this behavior
is fundamentally lacking.
Here we provide such microscopic insight by showing that the breaking of single bonds in the gel
network has consequences over large distances, in terms of cooperative particle rearrangements. We
do this by studying model restructuring gel networks, via numerical simulations, together with a new
space-resolved analysis of dynamical processes.
\begin{figure}
  \begin{center}
      \includegraphics[width=.9\columnwidth]{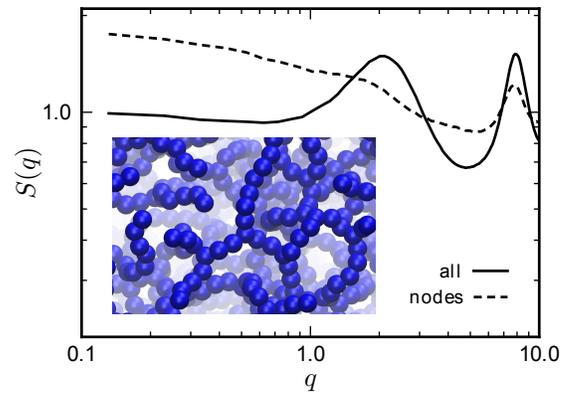}
  \end{center}
  \caption{Microstructure of the gel for $\rho=0.15$ and $\epsilon=20\, \kb T$. The static structure factor of the whole system
    (full line), and of the nodes only (dashed line) are depicted. Inset: real-space image of a
    section.\label{fig:structure}}
\end{figure}

Experiments revealed that the particle coordination in colloidal gels can be very low (2 to
3)~\cite{solomon} and that the bonds between particles can support significant
torques~\cite{pantina2005elasticity}, thus imparting local rigidity. Hence
we consider a minimal colloidal model 
with anisotropic interactions that stabilize open particle networks at low volume fraction
~\cite{delgado2007length,saw2009structural,more-models-1, *more-models-2, *more-models-3}.  For a microscopic particle configuration
$\{\mathbf{r}_{i}\}$ the interaction energy is:
\begin{equation*} 
U(\{\mathbf{r}_{i}\}) = \epsilon\left[\sum_{i > j}U_2(\vct{r}_{ij}/\sigma) + \sum_i\sum_{\substack{j>k}}^{j,k\ne i}U_3(\vct{r}_{ij}/\sigma,\vct{r}_{ik}/\sigma)\right]\,,
\end{equation*}
where $\vct{r}_{ij}$ is the vector connecting particles $i$ and $j$.  $U_2$ includes a repulsive
core and a narrow attractive well, whereas $U_3$ imposes an angular rigidity~\footnote{
  The two- and three-body terms in the interaction potential correspond to
  %\begin{equation*}
      $U_2(\vct{r})=A\left(a\,r^{-18}-r^{-16}\right)$
  %\end{equation*}
  and
  %\begin{equation*}
      $U_3(\vct{r},\vct{r}') = B\,\Lambda(r) \Lambda(r')\, 
      \exp\left[-\left(\frac{\vct{r}\cdot\vct{r}'}{rr'}-\cos\bar{\theta}\right)^2/w^2\right]\,,$
  %\end{equation*}
  respectively.
  The factor $\Lambda(r)=r^{-10} \left[1-(r/2)^{10}\right]^2 H(2-r)$, $H$ being the unit step
  function, ensures that only triplets of neighbors contribute to the energy.  Here we consider
  $A=6.27, a=0.85, B=67.27, \bar{\theta}=65^\circ, w=0.30$.}.
The parameters $\sigma$ and $\epsilon$ define respectively the units of length (equal to the
particle diameter) and energy: typical values for a colloidal system are $\sigma = 10-100\,\rm{nm}$
and $\epsilon = 1-100\,\kb T$, $\kb$ being Boltzmann constant and $T$ room
temperature~\cite{trappe_nature2001,luca_faraday2003,laurati_jor2011}. The unit of time is
$\sqrt{m\sigma^2/\epsilon}$, $m$ being the particle mass. We perform molecular
dynamics simulations~\cite{plimpton1995fast} of systems with $N=4000$ to $16384$ particles at number
density $\rho=0.15$, i.e., an approximate volume fraction $\phi \simeq 0.075$, with an interaction
strength $\epsilon = 20\, \kb T$ such that the particles self-assemble into a persistent spanning
network. Although very dilute, this model gel responds as a solid under mechanical loading, with a
finite elastic modulus~\cite{suppl}. The static structure factor $S(q) =
\frac{1}{N}\sum_{j,k}\exp[-i\vct{q}\cdot(\vct{r}_j-\vct{r}_k)]$ in Fig.~\ref{fig:structure}
quantifies spatial correlations between particle positions over distances $\simeq 2 \pi /q$ (where
$q$ is in units of $\sigma^{-1}$), regardless of whether they are connected or not through the
network. This simple model gel is made of chains linked by three-coordinated junctions or {\it nodes},
with a typical mesh size $l^*\simeq 3$-$4 \sigma$. 
$S(q)$ indicates the bonded particles ($q_b\simeq 8$
corresponds to distances of the order of the typical bond length) and the particles separated by
distances $\simeq l^*$ ($q^*\simeq 2$). 
The contribution to $S(q)$ of the network nodes
alone (Fig.~\ref{fig:structure}) indicates long range correlations throughout the network structure (low $q$) and 
the presence of clusters of nearby nodes (peak at $q_b$), suggesting that these
%\sout{the presence of densely connected regions.}
%that they 
are unevenly distributed in space.
We measure the density of nodes $\ci$ in the local environment around particle $i$ by counting their number within a
distance $l_{0}$ corresponding to $5$ bonds along the network.
%This is confirmed when we associate to each particle $i$ a connectivity index $c_i$ that measures the number of nodes lying within a distance of five bonds along the network: 
Its spatial distribution %of $\ci$ 
is indeed highly inhomogeneous, ranging from 0 in loosely connected
regions to $\approx 20$ in strongly connected ones. 
%The
%strong signal in the node $S(q)$ at small $q$ shows that there are long-range spatial correlations
%between nodes across loosely connected domains~\cite{delgado2010microscopic}.
% another common feature of dilute networks, which is that there are long-range spatial correlations between nodes across loosely connected domains \cite{delgado2010microscopic}.

The network {\it restructures} because thermal fluctuations favor breaking of existing bonds and formation of new ones. 
The fractions of two- and three- coordinated particles fluctuate about a value that is constant over our simulation time window, 
and that we use to define the topology of the network. To identify the contribution of the restructuring
processes to the dynamical properties of the gel, we compare the  restructuring networks to 
non-restructuring ones,   
%The networks {\it restructures} over time because thermal fluctuations favor breaking of existing
%bonds and formation of new ones. Hence the fractions of two- and three- coordinated particles
%fluctuate about a value that is roughly constant over our simulation time window, and that we use
%to define the topology of the network. We will now identify the contribution of restructuring
%processes to the dynamical properties of the gel by characterizing and comparing the microscopic
%dynamics of restructuring and non-restructuring networks. To do this we make it possible to
%constrain the restructuring networks 
by turning on a barrier in the interaction potential that
prevents bonds from breaking~\footnote{The barrier has the form ${U}_2^{\rm
    c}(\vct{r}_{ij}/\sigma)=C\exp\left[-(r_{ij}/\sigma-\gamma)^2/\delta^2\right]$, with $C=10.0$,
  $\gamma=1.2$ and $\delta=0.01$.}.

\begin{figure}
  \begin{center}
    \includegraphics[width=\columnwidth]{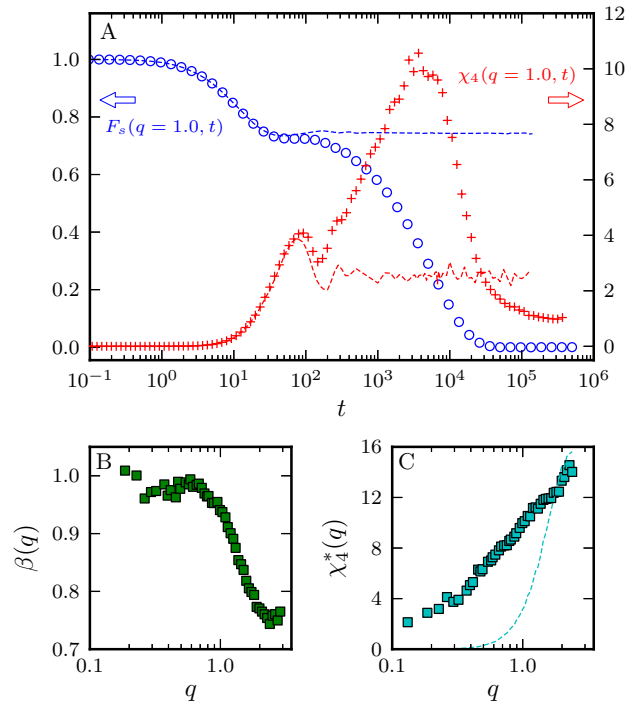}
  \end{center}
  \caption{(A): Incoherent scattering function $F_s(q,t)$ (left axis) and variance $\chi_{4}(q,t)$ (right axis) for $\rho=0.15$, $\epsilon=20\,\kb T$ and $q=1.0$, plotted as a function of time, for the
    restructuring (symbols) and constrained (dashed line) networks; (B): $q$-dependence of the
    stretching exponent $\beta$; (C): Maximum value $\chi_{4}^{*}(q)$ of $\chi_{4}(q,t)$ over time
    plotted as a function of $q$ for the restructuring (symbols) and constrained (dashed line)
    networks.
    \label{fig:sisf}}
\end{figure}

\begin{figure}
  \begin{center}
    \includegraphics[width=1.0\columnwidth]{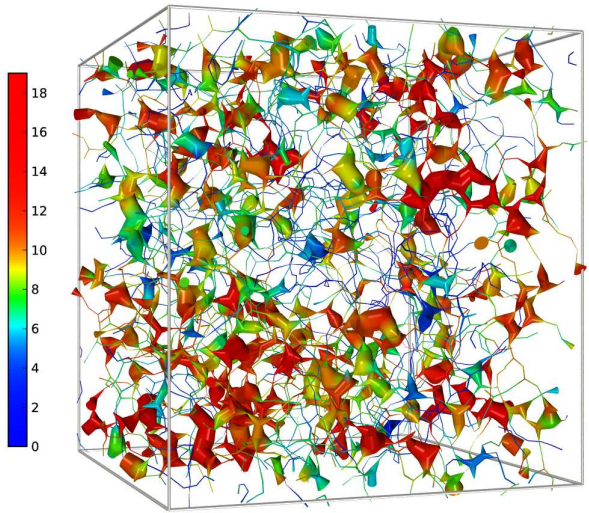}
  \end{center}
  \caption{Spatial map of propensity for bond-breaking $b_i$ relative to a time lag $\tau=10^3$ and
    a given configuration $\mathscr{C}_0$. The thickness of the segments representing inter-particle
    bonds is proportional to the propensity for bond-breaking of the bonded particles. The color
    code represents the local density of nodes $\ci$.
    \label{fig:isoconf_bond}}
\end{figure}
\begin{figure}
  \begin{center}
\includegraphics[width=1.0\columnwidth]{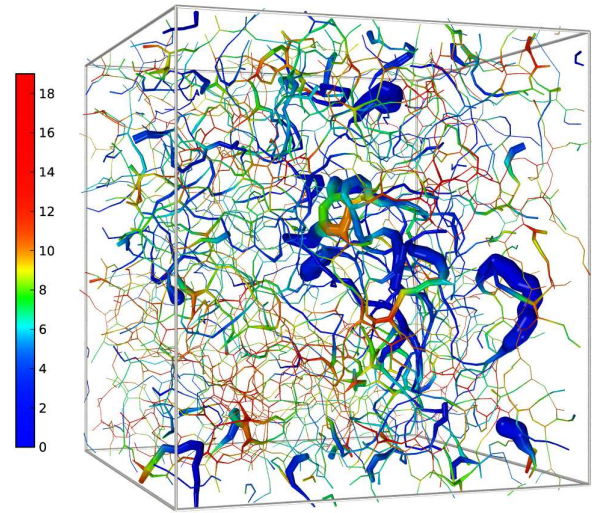}
  \end{center}
  \caption{Spatial map of propensity for displacement $p_i$ relative to to the same configuration
    and time lag as Fig.~\ref{fig:isoconf_bond}. Segments represent interparticle bonds and their
    thickness is proportional to the propensity for displacement of the bonded particles. The
    color code represents the local density of nodes $\ci$.
    \label{fig:isoconf}}
\end{figure}

We analyse how the structure changes via the mean degree of self-correlation in
particle positions over distances $\simeq 2\pi/q$ and over time $t$, as measured by the incoherent
scattering function $F_s(q,t) = \left\langle\Phi_{s}(q,t)\right\rangle$, where $\langle ... \rangle$
indicates a time average and $\Phi_s(q,t) =
\frac{1}{N}\sum_j\exp\left[-i\vct{q}\cdot\left(\vct{r}_j(t)-\vct{r}_j(0)\right)\right]$.
%%WC%%\begin{equation*}
%%WC%%\Phi_s(q,t) = \frac{1}{N}\sum_j\exp\left[-i\vct{q}\cdot\left(\vct{r}_j(t)-\vct{r}_j(0)\right)\right].
%%WC%%\end{equation*} 
Our data show a strong dependence on the length scale sampled by $q$ and a complex time decay of
correlations for wave vectors $q \le q^{*}\simeq 2$. Fig.~\ref{fig:sisf}A shows $F_s(q,t)$ at
$q=1.0$ (left axis): the two-step decay with a plateau suggests that particles undergo a transient
localization process similar to caging in dense glassy suspensions~\cite{glasses}. The decays in the
restructuring network (symbols) and the constrained one (dashed line) overlap up to $t_{\rm loc}
\simeq 100$. This indicates that the initial decay of correlations is mainly due to motion that is
present also in the constrained network and that the plateau value (related to the particle
localization) is determined by fixing the network topology (i.e., the mean fractions of $2$- and
$3$- coordinated particles present). The final decay of correlations after the plateau is instead
due to the network restructuring. The decay from the plateau is well fitted by a
stretched-exponential, $F_s(q,t)\sim\exp[-(t/\tau_{q})^\beta]$, with $\beta (q)<1$ and decreasing to
$\sim 0.7$ with increasing $q$ (Fig.~\ref{fig:sisf}B). This qualitative change in the long-time
decay, upon varying $q$, suggests that qualitatively different relaxation processes must come into
play over different length-scales. The shape of the decay for $q \leq 2.0$ 
is typically associated with cooperative dynamics in glassy systems \cite{dh}.
We characterize this in the gel in a spatially and temporally
averaged way using the variance
$\chi_4(q,t)=N[\langle|\Phi_s(q,t)|^2\rangle-\langle\Phi_s(q,t)\rangle^2]$,
%%WC%%\begin{equation*} 
%%WC%%\chi_4(q,t)=N[\langle|\Phi_s(q,t)|^2\rangle-\langle\Phi_s(q,t)\rangle^2]\,,
%%WC%%\end{equation*} 
that quantifies dynamical heterogeneity due to cooperativity. 
This dynamic susceptibility detects fluctuations from the
mean degree of correlation in single particle displacements due to spatial correlations of the
dynamics, i.e., to particles undergoing cooperative motion over time $t$ and distance $\simeq
2\pi/q$ \cite{dh1,dh3,dh}. Fig.~\ref{fig:sisf}A shows $\chi_4(q,t)$ (right axis) for the restructuring
(symbols) and constrained networks (dashed line) at $q=1.0$. The curves coincide up to $t \simeq
t_{\rm loc}$ and therefore we can ascribe the initial increase of $\chi_4(q,t)$ to vibrational
motion present in both networks.  At longer times, in the constrained network $\chi_4(q,t)$ remains
constant around its asymptotic value \cite{abete_pre2008,fierro_jstat2008}. In contrast, in the
restructuring network $\chi_4(q,t)$ develops a significant peak at $t > t_{\rm loc}$ and decays to
the asymptotic value of $1$ on the same timescale as $F_s(q,t)$ decays to 0, again very similar to
what found in dense glassy suspensions \cite{dh}. The maximum value of $\chi_4(q,t)$ over time,
$\chi_{4}^{*}(q)$, increases with the number of particles undergoing cooperative dynamics over
length-scales $\simeq 2\pi/q$ \cite{dh2}.  Fig.~\ref{fig:sisf}C shows that, for all $q \le q^*
\simeq 2$, it is significantly higher in the restructuring network. Hence the long-time 
restructuring is cooperative over this range of length-scales.

In order to elucidate how such cooperative processes take place in the gel, we perform a
spatially-resolved analysis of the dynamics discussed so far.  Let $\mathscr{C}_0=\{\vct{r}_i^0\}$
denote the configuration of the gel, i.e. the set of all particle positions, at time $t=0$, and let
$\mathscr{C}_{\tau}=\{\vct{r}_i^\tau\}$ be the corresponding configuration after an elapsed time
interval $\tau$ during which the network was able to restructure. The set of particle displacements
$\{\vct{r}_i^{\tau}- \vct{r}_i^{0} \}$ is determined not only by the restructuring process, but also
by the fast vibrational dynamics. With this in mind we define the sets of \emph{intrinsic} particle
positions $\{\vct{\hat{r}}_{i}^0\}$ and $\{\vct{\hat{r}}_{i}^\tau\}$ by constraining the bonds of
the corresponding network configuration ($\mathscr{C}_0$ and $\mathscr{C}_\tau$, respectively) and
computing the temporally averaged position of each particle over a sufficiently long time
window. The intrinsic displacements $\{\vct{\hat{r}}_i^{\tau}-{\vct{\hat{r}}}_i^0\}$ quantify the
net effect of restructuring, free from any significant contributions of the network
vibrations~\cite{suppl}.

We iterate the mapping $\mathscr{C}_0\rightarrow\mathscr{C}_\tau$ starting from the same network
configuration $\mathscr{C}_0$ and using different sets of initial particle velocities drawn from a
Maxwell-Boltzmann distribution at the same prescribed temperature $T$.  This iso-configurational
(IC) ensemble of trajectories allows us to identify, for each particle, its tendency to contribute
to specific dynamical processes over a time
$\tau$ in a way that can be directly related to the microstructure
~\cite{widmer2004reproducible,widmer_JCP2007,asaph_nphys,widmer_JCP2009}.  Since we are
interested in processes that are relevant to the network restructuring we use $\tau\simeq10^3$,
which corresponds to the time $t$ where $F_{s}(q^{*},t)$ has decayed to $\sim1/2$ of its value at
the plateau (and where $\chi_4(q^{*},\tau) \simeq \chi_4^{*}(q^{*})$).

First we use this approach to identify the parts of the structure where bond-breaking, which is
responsible for network restructuring, is more likely to occur.  This is achieved by defining for
each particle $i$ a propensity for bond-breaking $b_{i}=\langle n_{i} \rangle_{\mathrm{IC}}$, where
$n_{i}=1$ if the particle looses, along one of the trajectories, any of the particles bound to it,
and 0 otherwise; the angular brackets denote an average over the IC ensemble of trajectories.
Fig.~\ref{fig:isoconf_bond} shows the spatial distribution of the propensity for bond-breaking for a
typical initial configuration: bonded particles are represented by segments of thickness
proportional to $b_{i}$ and colored according to the local density of nodes $\ci$, from dark blue
(low $\ci$) to red (high $\ci$). The figure shows that quite unexpectedly
bond-breaking events responsible for network restructuring tend to take place in regions with high
local density of nodes. We have computed the
  spatial distribution of local stresses and find that bonds in regions with a higher 
local density of nodes are indeed subjected to higher-than-average tensile stress~\cite{suppl}.

%% The spatial distribution of local stresses in the gel network~\cite{suppl} suggests that this is a
%% consequence of the fact that, in regions with higher local density of nodes, bonds are subjected to
%% higher-than-average tensile stress.

The distribution of intrinsic displacements over our IC ensemble (at the time $\tau = 10^3$) 
quantifies for each particle its tendency to undergo a significant displacement as a consequence 
of bond-breaking, thus contributing to the cooperative network restructuring.  The second moment
$p_i=\left\langle(\vct{\hat{r}}_i^\tau-\vct{\hat{r}}_i^0)^2\right\rangle_{\mathrm{IC}}$, which
we will call the propensity for displacement of particle $i$ in $\mathscr{C}_0$, provides a spatial map of
the effects of bond-breaking events in terms of intrinsic particle displacements.
%displacements responsible for the cooperative dynamics detected during the restructuring 
%of the network (i.e.,  for the non-exponential long-time decay of correlations in the gel).
Fig.~\ref{fig:isoconf} shows the spatial distribution of the propensity for displacement for the
same initial configuration as in Fig.~\ref{fig:isoconf_bond}. The thickness of the segments now
indicates in which parts of the network the significant displacements responsible for the
cooperative restructuring dynamics (Fig.~\ref{fig:sisf}) tend to take place. These large intrinsic displacements occur in regions with low local density
of nodes, i.e., far away from the bond-breaking events that caused them. We find that these are also
the parts of the network more prone to undergo non-affine displacements due to the build-up of
internal stresses when the gel is deformed~\cite{suppl}. The cooperative processes we have uncovered
are therefore also important for the mechanical response of the gel.

\begin{figure}
  \begin{center}
    \includegraphics[width=.9\columnwidth]{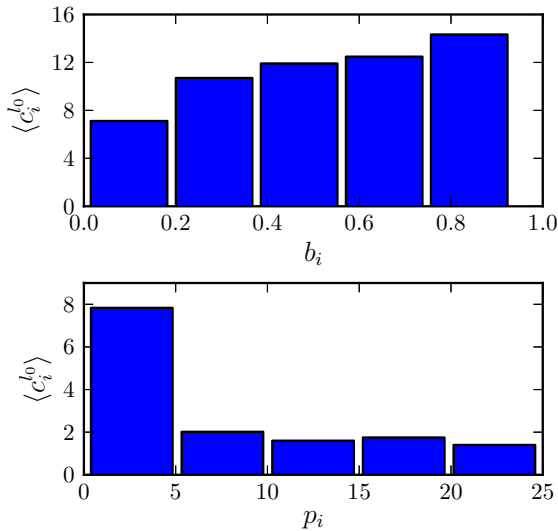}
 \end{center}
 \caption{Average local density of nodes $\langle \ci \rangle$ as a function of propensity
   for bond-breaking $b_i$ (top) and as a function of propensity for displacement $p_i$
   (bottom).\label{fig:hist}}
\end{figure}

We have quantified these observations by analyzing data obtained from $6$ independent initial
configurations of a network with $4000$ particles, resulting in a set of $24000$ triplets $\{\ci,
b_i, p_i\}$~\cite{suppl}.~In Fig.~\ref{fig:hist} we have sorted the particles into five groups
according to the values of $b_i$ (top) and $p_i$ (bottom), and plotted for each group the average
local density of nodes $\ci$.~Quantifying the degree of linear correlation using Pearson's
product-moment correlation coefficient~\cite{rodgers1988thirteen}, we find indeed a negative
correlation between $p_i$ and $\ci$ (Pearson's product-moment correlation coefficient $r \simeq
-0.35$) and a positive correlation between $b_i$ and $\ci$ ($r \simeq 0.42$).

%\section{Discussion}

The picture that emerges is that bond-breaking tends to take place in regions of the network that
are more locally connected and experience higher local stresses. Bond-breaking then induces large
rearrangements further away, i.e., over length-scales well beyond the network mesh size.  These
cooperative particle displacements ultimately lead to the network restructuring, characterized by
the complex decay of time correlations in the relaxation dynamics
\cite{duri2006,luca_prl2009,maccarrone}. 
%A further indication of the non-local coupling between
%structure and dynamics is that there are many particles with low (high) $\ci$ that do not have high
%$p_i$ ($b_i$), i.e. the probability of local network restructuring does not appear to be determined
%by the local structure alone. The reason for this is likely that different parts of the network are
%able to talk to one another, via the links and nodes, resulting in non-local constraints and
%feedback. 
This feedback between local and collective processes,
%Such feedback between local and collective processes is strongly 
reminiscent of the coupling between particle localization and overall structural relaxation in the dynamics of dense
glassy suspensions~\cite{asaph-candelier_prl, leporini2011},  
%Having recognized that here such
%feedback originates from the long-range correlation entailed in the network structure, 
explains the similarities between dilute gels and dense glasses in the caging processes, two-step decay of
time correlations with stretched exponential behavior, and dynamical heterogeneity. 

The microscopic picture we obtain is that the gel contains weaker regions (where local stresses are higher and
breaking is more likely to occur), which is consistent with arguments made to rationalize the mechanical
response of restructuring gels in experiments \cite{laurati_jor2011,pochan_softmatter2010}. We  
%results, however, 
show, however, that the consequences of bond-breaking are non-local, cooperative
displacements involving parts of the structure that are relatively far away, in regions where
non-affine displacements are more likely to concentrate under deformation. Hence a
significantly more complex scenario emerges for the mechanical response, where the mesoscale organization of
the gel network has a significant role. Our results suggest that under deformation higher tensile stresses are 
likely to build up, and favor bond-breaking, in densely connected regions. Moreover, non-affine displacements 
under deformation are related to the irreversible rearrangements responsible for local plastic events 
\cite{bocquet_nature,lemaitre2007plastic}
that hinder relaxation of local stresses upon unloading the material. 
%Hence our findings indicate how the 
%amount and rate of relaxation of local stresses in different parts of the structure will be ultimately controlled 
%also by the environment relatively far away along the network. 
This microscopic, quantitative information, available from our study, on when and where local irreversible events are 
more likely to occur, can now be incorporated into non-local constitutive models for the viscoelastic response of 
colloidal gels and similar materials.

This work was supported by Swiss National Science Foundation (SNSF) (Grants
No. PP00P2\_126483/1 and No. IZK0Z2\_141601). A.W. also thanks the Australian Research Council for support.

\makeatletter 
\renewcommand{\thefigure}{S\@arabic\c@figure}
\makeatother
\setcounter{figure}{0}

\section*{Supplementary Information}
\subsection{S1. Mechanical tests}

\begin{figure}
\begin{center}
  \includegraphics[width=1.\columnwidth]{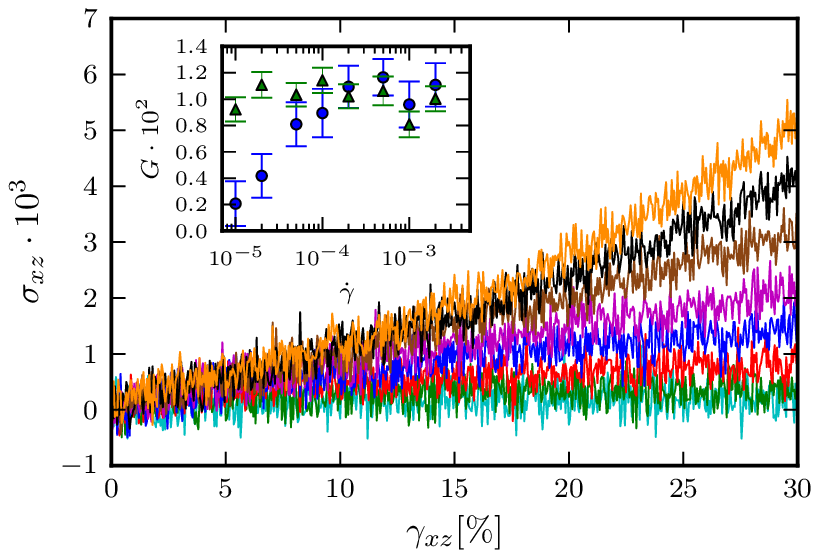}
  \caption{Main plot: shear stress $\sigma_{xz}$ as a function of shear strain $\gamma_{xz}$ for
    different values of the shear rate $\dot{\gamma}$ (from top to bottom: $2\cdot 10^{-3}$,
    $10^{-3}$, $5\cdot 10^{-4}$, $2\cdot 10^{-4}$, $10^{-4}$, $5\cdot 10^{-5}$, $2\cdot 10^{-5}$,
    $10^{-5}$). Inset: shear modulus as a function of shear rate, obtained by fitting a linear
    function to the stress-strain curves in the region of small deformation $0 \le \gamma_{xz} \le
    5\%$ (circles); shear modulus of the non-restructuring network (triangles).\label{fig:shear}}

\end{center}
\end{figure}

We have carried out non-equilibrium molecular dynamics simulations of a restructuring gel network
with $N=4000$ particles, density $\rho=0.15$ and an interaction strength $\epsilon = 20\,\kb T$
(reduced temperature $T=0.05$). The network was sheared in the $xz$ plane with a constant shear rate
$\dot{\gamma}$ up to a final strain $\gamma_{xz}=30\%$, and the shear stress $\sigma_{xz}$ was
recorded. The tests were performed using SLLOD equations of motion~\cite{evans2007statistical}
implemented in LAMMPS~\cite{plimpton1995fast}, and the temperature was controlled by means of a
chain of Nos\'e-Hoover thermostats.  We show results obtained by averaging over a set of $100$
independent initial configurations of the network.

The stress-strain curves corresponding to a set of shear rates spanning more than two orders of
magnitude -- from $10^{-5}$ to $2\cdot 10^{-3}$ -- are presented in Fig.~\ref{fig:shear}. Although
for large deformations the response of the network depends markedly on the shear rate, there exists
a linear elastic regime extending approximately up to $\gamma_{xz}\approx 5\%$ from which we can
extract a small, but finite shear modulus $G\approx 0.01\, \epsilon/\sigma^3$ for $\dot{\gamma} \ge
5\cdot 10^{-5}$ (blue circles in the figure inset). Hence our gel network has a solid-like elastic
response at relatively low shear rate.  In a typical colloidal system characterized by $\epsilon =
50\,\kb T$, $\sigma=100$ nm, this corresponds to a shear modulus $G\approx 2$ Pa, which is
consistent with what measured in experiments on dilute colloidal gel networks
\cite{trappe_nature2001,luca_faraday2003}. For even lower shear rates the modulus drops towards
zero, presumably due to thermally activated restructuring of the network. This hypothesis is
supported by the comparison with a non-restructuring network, whose shear modulus is approximately
constant over the whole range of shear rates investigated (green triangles in the figure inset).

\subsection{S2. On the propensity for displacement}

\begin{figure}
\begin{center}
  \includegraphics[width=.9\columnwidth]{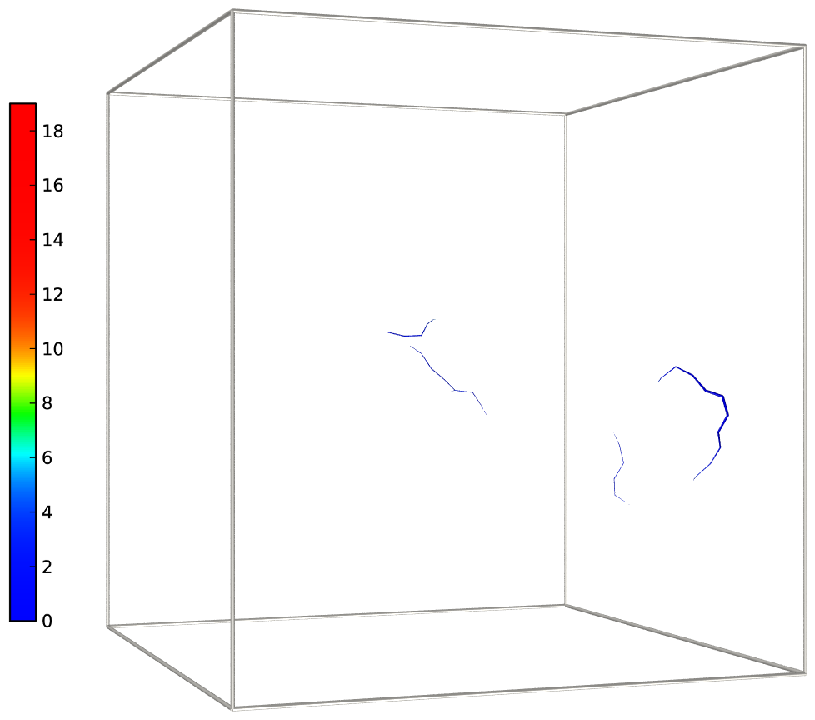}
  \caption{Spatial map of propensity for non-vibrational displacement for the
    same initial configuration as Figs.~3 and 4 of the main text, but where
    the bonds have been constrained to not break. Segments represent
    interparticle bonds and their thickness is proportional to the propensity
    for displacement of the bonded particles; the scale is the same as in
    Fig.~4 of the main text. The color code represents the local density of
    nodes.\label{fig:pi_cnstr}}
\end{center}
\end{figure}

The iso-configurational analysis was originally introduced for dense glassy
alloys~\cite{widmer2004reproducible} and needs to be modified for low-density materials. In our low
density gel network the amplitude of particle vibrations at fixed inter-particle bonds is far bigger
than the amplitude of cage rattling in a typical dense system. In order to study network relaxation
one needs a way to filter out these vibrations from the underlying displacements due to
restructuring.  As described in the main text, we have accomplished this task by associating to any
given network configuration an \emph{intrinsic} configuration, that we compute by constraining the
inter-particle bonds and evaluating the average position of each particle during a simulation at
constant temperature. Since bond-breaking cannot take place, the time evolution samples the
vibrational dynamics of the given configuration. We compute the intrinsic positions by averaging
over a time window $\Delta t_{\rm av}=2.5\cdot 10^4$ in reduced units, which is sufficiently long
for most structural correlations to decay to zero in the unconstrained system.

As a demonstration that this approach is able to filter out most of the displacements that are not
due to network restructuring, we show in Fig.~\ref{fig:pi_cnstr} the result of the
iso-configurational analysis performed on a \emph{non-restructuring} network. The network
configuration is the same as the one shown in Figs.~3 and 4 of the main text, and the time lag for
the isoconfigurational trajectories is likewise $\tau=10^3$. Ideally, one would expect each
particle to display zero propensity for displacement. In practice, a subset of the particles show a
small but non-zero propensity: this is due to very slow network rearrangements that do not entail
breaking of existing bonds or formation of new bonds, but still affect the average position of some
of the particles. For instance, a chain initially confined to a region of space by the surrounding
chains can break free of its cage and start to oscillate in a different region: this would affect
the average position of the particles in the chain, even though no restructuring -- that is no
change of inter-particle bonds -- has taken place. The contribution of processes of this latter kind
to the propensity, however, are very small compared to the displacements due to restructuring. By
comparing the propensity for displacement in the restructuring network, $p_i$, to the same quantity
evaluated in the constrained network, $p_i^c$, one can define a noise-to-signal ratio as
\begin{equation*}
  {\rm NSR}_p = \frac{1}{N} \sum_{i=1}^{N} \frac{p_i^c}{p_i}\,.
\end{equation*}
For the configuration shown in Fig.~\ref{fig:pi_cnstr} the ${\rm NSR}_p < 1\%$, i.e. the displacements
not directly ascribable to network restructuring contribute on average less than $1\%$ of the
total per-particle signal.

\subsection{S3. Correlation between propensities and local density of nodes}

\begin{figure}
\begin{center}
  \includegraphics[width=.9\columnwidth]{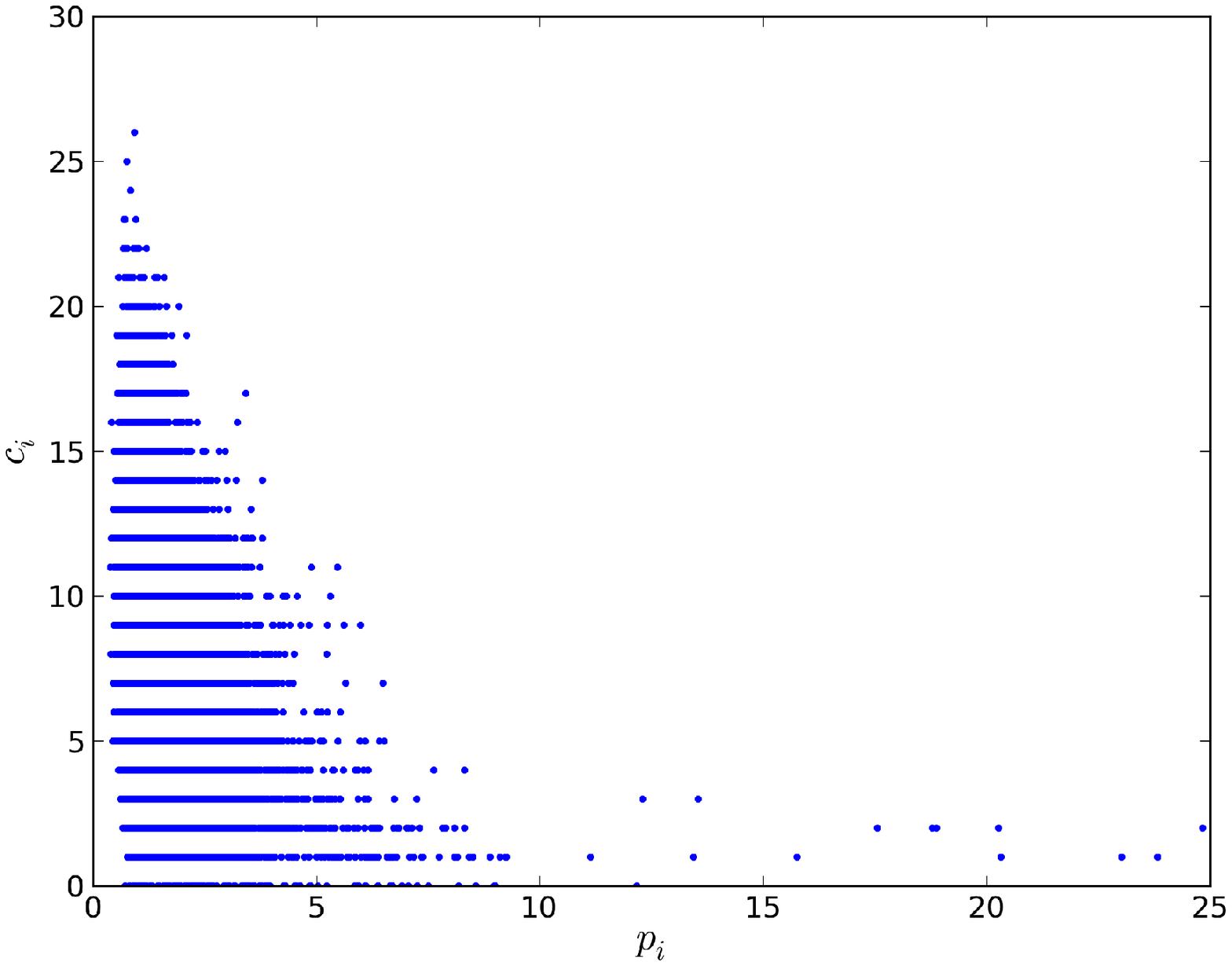}
  \caption{Scatter plot $\ci$ vs. $p_i$.\label{fig:pi_sp}}
\end{center}
\end{figure}

\begin{figure}
\begin{center}
  \includegraphics[width=.9\columnwidth]{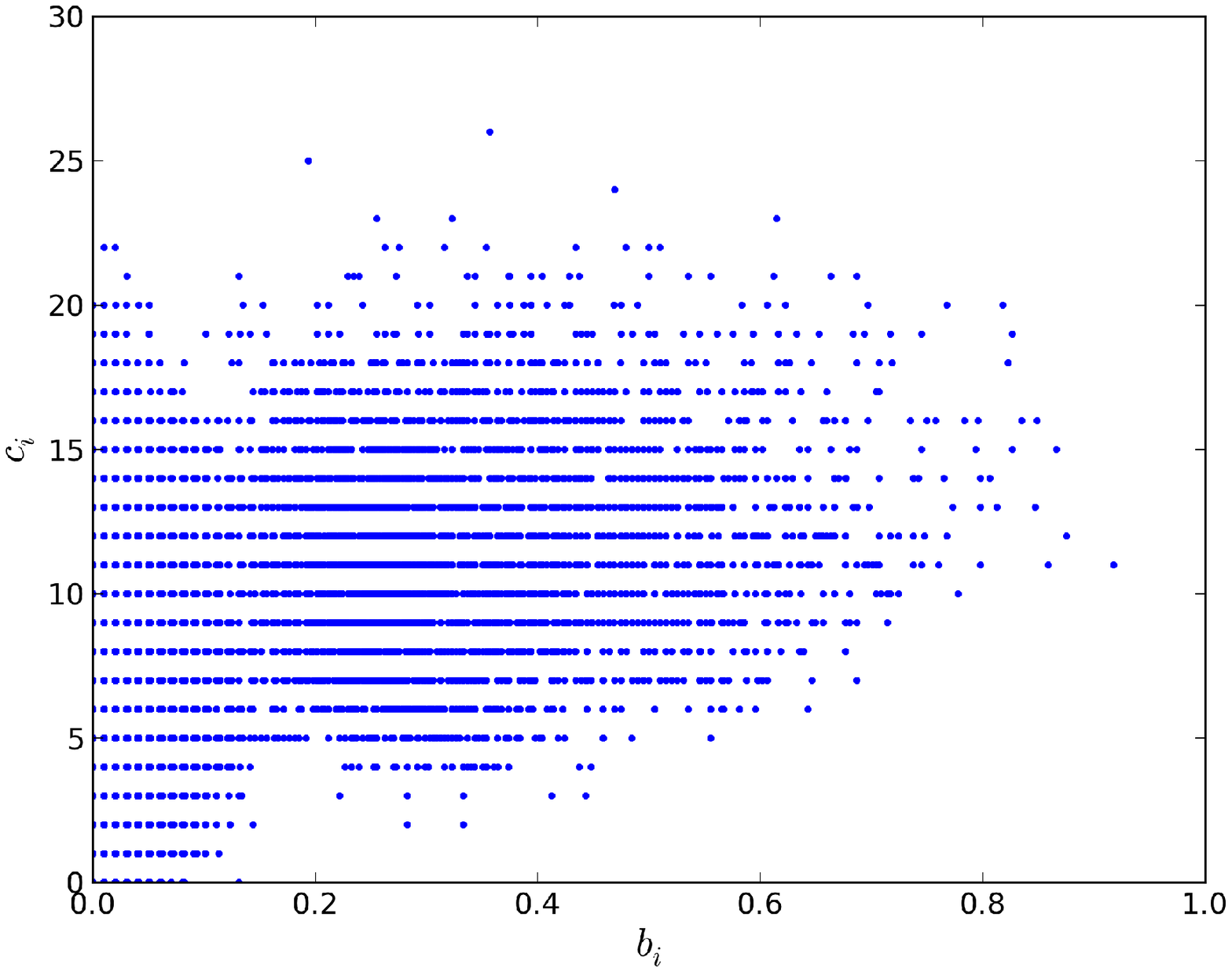}
  \caption{Scatter plot $\ci$ vs. $b_i$.\label{fig:bi_sp}}
\end{center}
\end{figure}

In performing the iso-configurational analysis we have considered six independent configurations of a
network with $N=4000$ particles; for each configuration, an ensemble of 100 trajectories has been
generated. This allowed us to associate to each particle $i$ a local density of nodes $\ci$ --
a topological attribute of the particle in the initial configuration -- as well as the propensities
for non-vibrational displacement ($p_i$) and bond breaking ($b_i$), which are dynamical properties
calculated by averaging over the ensemble of trajectories. In this way we gathered a set of 24000
triplets $\{\ci, p_i,b_i\}$.  The raw scatter plots $\ci$ vs. $p_i$ and $\ci$
vs. $b_i$ are shown in Figs.~\ref{fig:pi_sp} and~\ref{fig:bi_sp}, respectively. It is apparent that
particles characterized by a high propensity for displacement tend to be in regions with low density
of nodes; in contrast, particles characterized by a high propensity for bond breaking tend to
localize in regions with high density of nodes. What is also apparent is that the inverse
relationship does not hold, i.e. particles with low (high) $\ci$ often do not have high $p_i$
($b_i$). This indicates that the local particle environment is not the only structural property that
affects where relaxation takes place.

\subsection{S4. Stress distribution}

Bond breaking happens preferentially in regions with a high density of nodes (i.e. crosslinks). In
order to ascertain whether this is due to a concentration of stresses in those regions, we have
calculated the stress distribution in the gel.

Local stresses are obtained by dividing the simulation box into a set of $n_c$ cubic cells,
$\{\mathcal{C}_m\}_{m=1\dots n_c}$, and associating with each cell an average stress tensor
$\sigma_{\alpha\beta}^m=\langle s_{\alpha\beta}^m\rangle$, where $s_{\alpha\beta}^m$ is the
instantaneous stress tensor corresponding to a specific particle configuration, and the angular
brackets denote a time average. The instantaneous stress tensor for each cell is computed according
to the standard virial formula~\cite{2009plimpton} in the following way:
\begin{equation}
  s_{\alpha\beta}^m = -\frac{1}{V_m}\sum_{i\in\mathcal{C}_m} w_{\alpha\beta}^i\,,
\end{equation}
where $V_m$ is the cell volume, the sum runs over the particles contained in the cell, and
$w_{\alpha\beta}^i$ represents the contribution to the stress tensor of the interactions involving
particle $i$. The latter is defined by splitting the contribution of each interaction evenly among
the particles that participate in it:
\begin{equation}
\begin{aligned}
w_{\alpha\beta}^i &= \frac{1}{2}\sum_{n=1}^{N_2} (r^i_{\alpha}F^i_{\beta} + r'_{\alpha}F'_{\beta}) +\\
&\frac{1}{3} \sum_{n=1}^{N_3} (r^i_{\alpha}F^i_{\beta} + r'_{\alpha}F'_{\beta} + r''_{\alpha}F''_{\beta})\,.
\end{aligned}
\end{equation}
In the previous expression the first sum runs over the $N_2$ pair interactions that particle $i$ is
part of, $\mathbf{r_i}$ and $\mathbf{r'}$ are the positions of the two interacting particles, and
$\mathbf{F_i}$ and $\mathbf{F'}$ are the forces on the two particles resulting from the
interaction. Along the same lines the second sum takes into account the $N_3$ three-body
interactions involving particle $i$. \footnote{The ``particle stress'' $w_{\alpha\beta}^i$ is
  readily available in \texttt{LAMMPS} thanks to the fix \texttt{stress/atom}.} 
%% As a last remark, we
%% note that the classical definition of the stress tensor would also contain a kinetic term related to
%% particles entering or leaving the cells; however, since we interested here in average tension of the
%% bonds, we consider only the part due to inter-particle interactions.

We have applied the analysis just outlined to a gel network with $N=16384$ particles, density
$\rho=0.15$ and interaction strength $\epsilon=20\,\kb T$ (reduced temperature $T=0.05$). The
simulation box has been divided in $n_c=7^3$ cells, so that each cell contains on average around
$50$ particles. The time window for the temporal averaging has been chosen as $\Delta t=250$, which
is long enough to account for the vibrational dynamics of the network, but yet short enough for the
network structure not to be significantly perturbed by the restructuring process (during the
prescribed time window less than 5\% of the particles change neighbors). For each cell we have also
computed the average number density of nodes $\rho_{\rm nodes}^m$, a quantity that measures the
local concentration of crosslinks.

\begin{figure}
\begin{center}
  \includegraphics[width=.9\columnwidth]{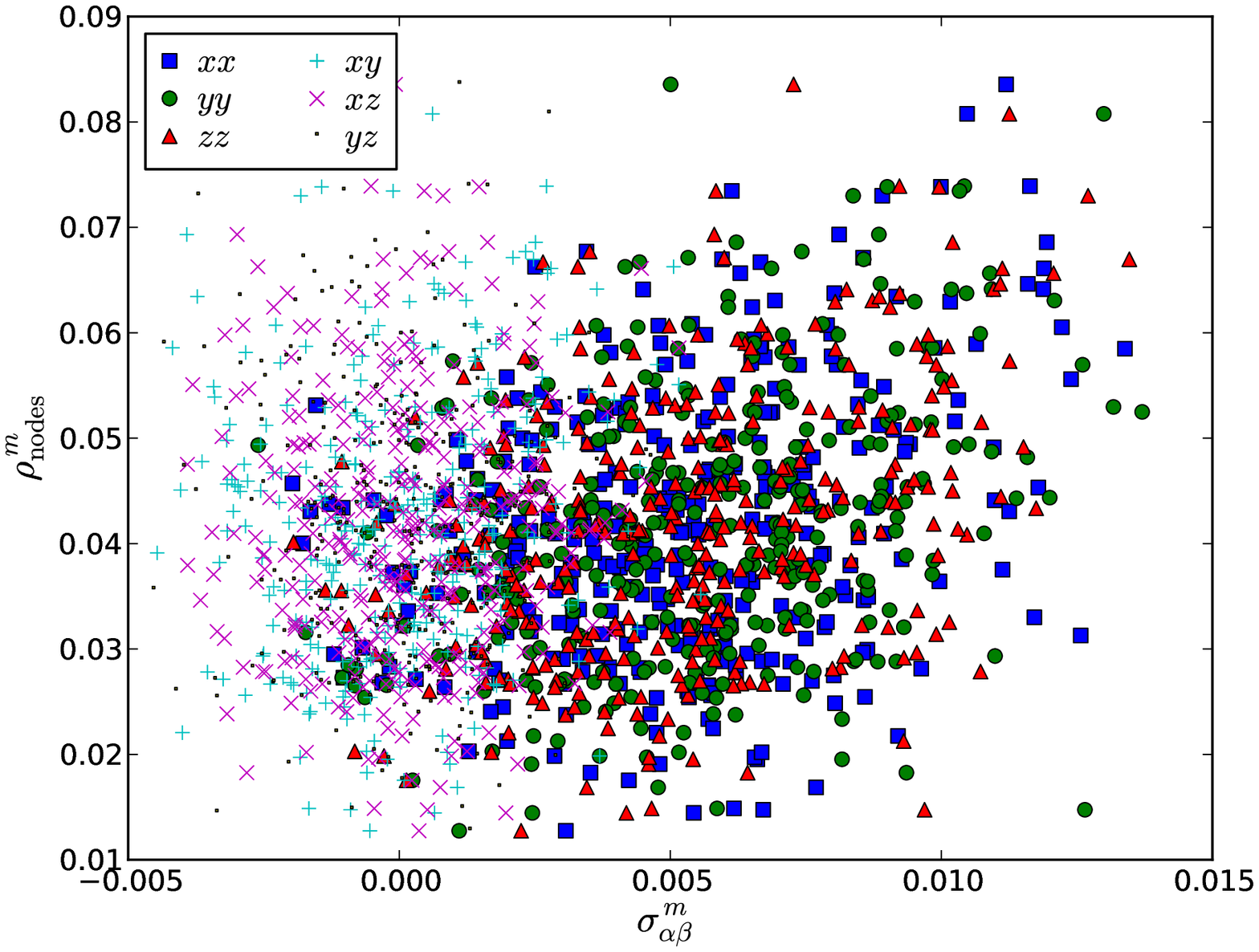}
  \caption{Scatter plot of the components of the local stress tensor $\sigma^m_{\alpha\beta}$
    vs. the local density of nodes $\rho^m_{\rm nodes}$. \label{fig:st_vs_rhonodes}}
\end{center}
\end{figure}

In Fig.~\ref{fig:st_vs_rhonodes} we show the scatter plot of the components of the local stress
tensor vs. the local density of nodes. The shear components $xy$, $xz$ and $yz$ are on average zero
and do not show any particular correlation with the density of nodes. On the contrary, the diagonal
components $xx$, $yy$ and $zz$ are greater than zero and have a positive correlation with the
density of nodes. This means that the cells are under isotropic tension, the tension being greater
where the density of nodes is higher.

\begin{figure}
\begin{center}
  \includegraphics[width=.9\columnwidth]{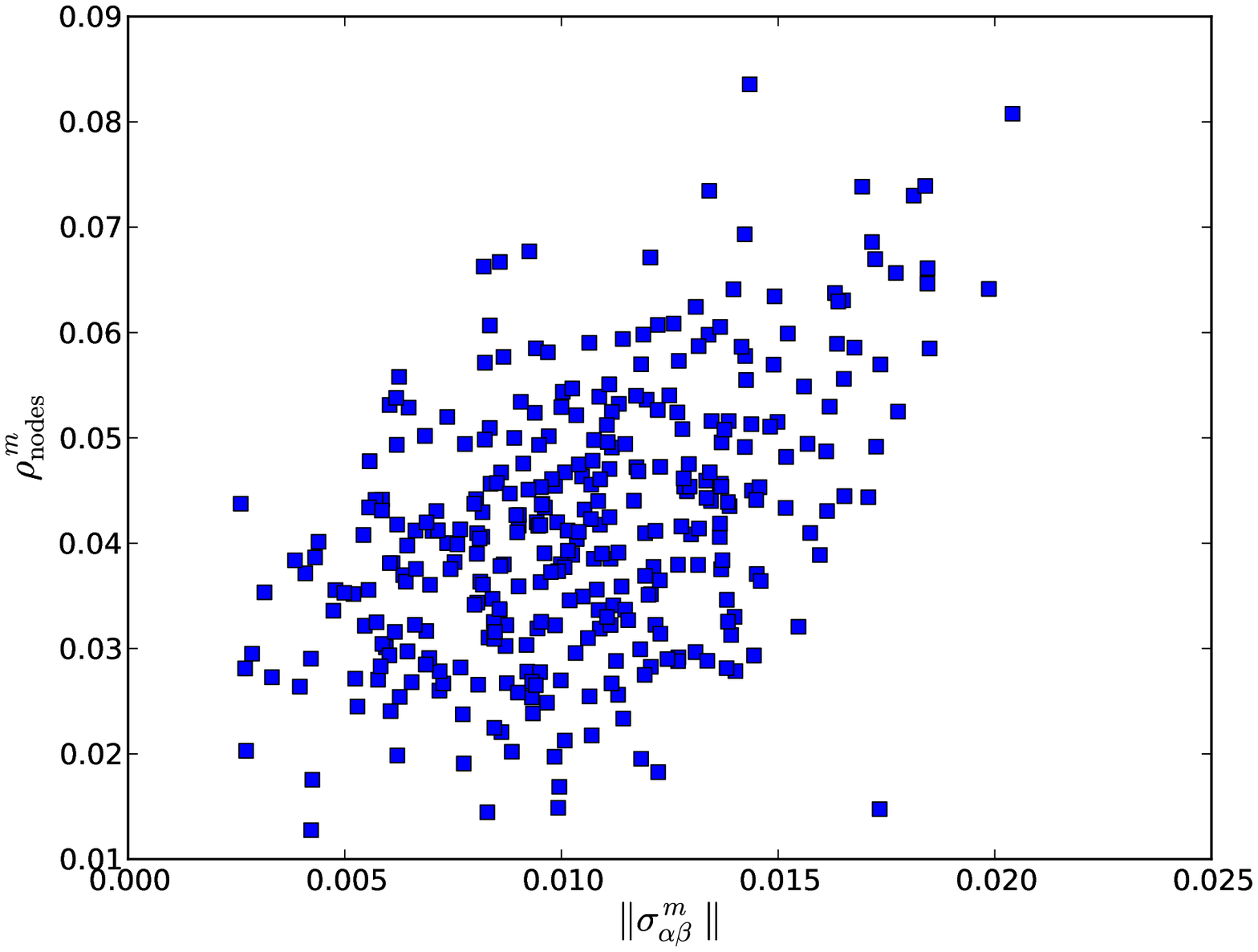}
  \caption{Scatter plot of the norm of the local stress tensor $\Vert\sigma^m_{\alpha\beta}\Vert$
    vs. the local density of nodes $\rho^m_{\rm nodes}$.\label{fig:stnorm_vs_rhonodes}}
\end{center}
\end{figure}

The same trend is evinced from Fig.~\ref{fig:stnorm_vs_rhonodes}, where the norm of the local stress
tensor is plotted vs. the local density of nodes. The positive correlation between the two
quantities is assessed by a Pearson's coefficient $r=0.47$.

\begin{figure}
\begin{center}
  \includegraphics[width=.9\columnwidth]{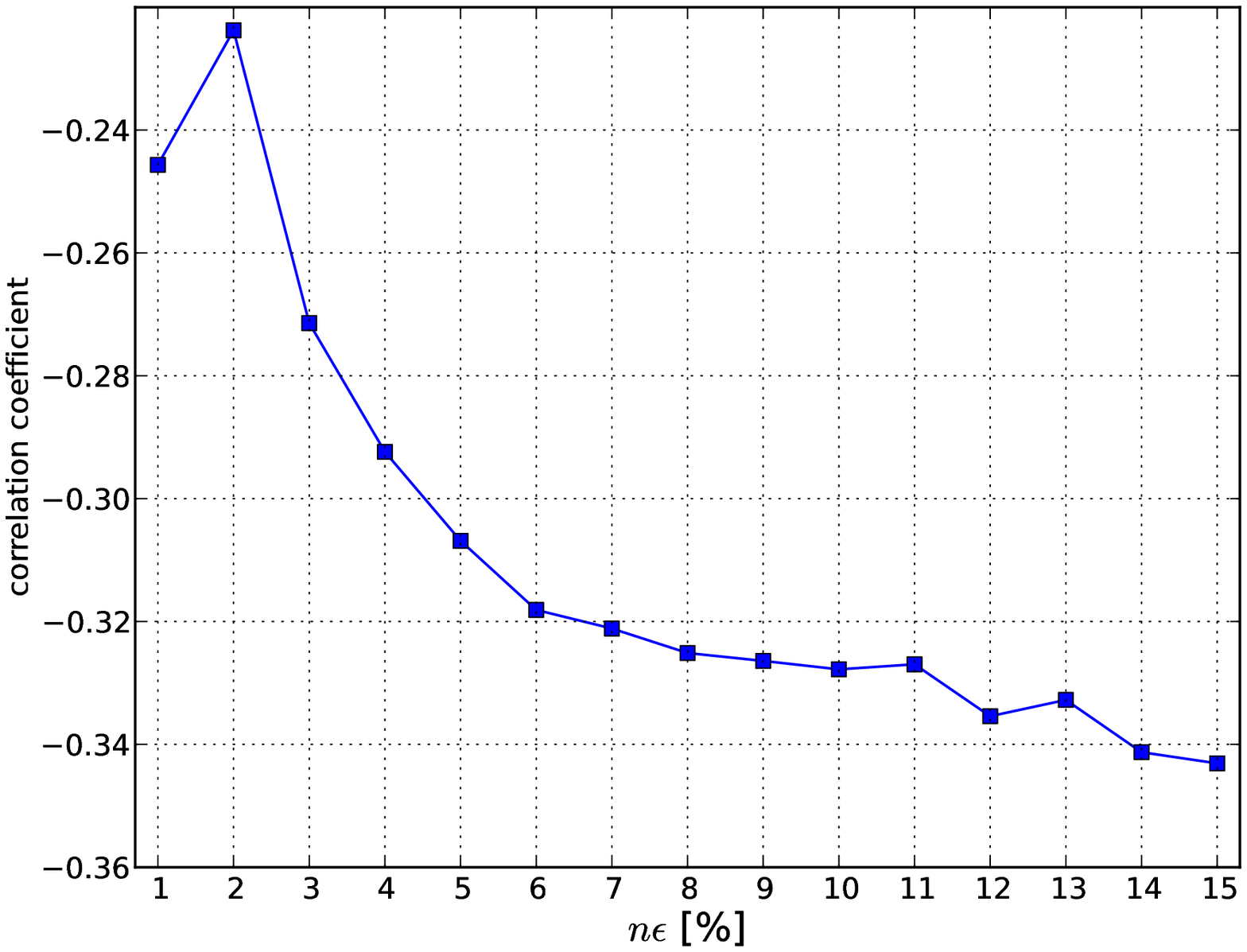}
  \caption{Pearson's coefficient of correlation between the norm of the non-affine displacements and
    the local density of nodes, $r(\{\Vert\mathbf{d}_i\Vert\}_{n\varepsilon}, \ci)$, plotted as a
    function of the strain $n\varepsilon$. \label{fig:nad_corr}}
\end{center}
\end{figure}

\subsection{S5. Non-affine displacements}

In order to understand which parts of the gel are more prone to displacement as a consequence of
internal stresses we have performed a quasi-equilibrium deformation of the network at zero
temperature and looked at the non-affine displacement field. The procedure is as follows. 

A configuration of the network at finite temperature $\mathscr{C}_0$ is first relaxed to the closest
minimum in the potential energy landscape $\tilde{\mathscr{C}}_0$, i.e. the closest ``inherent
structure'', by slowly draining energy from the system by means of a fictitious frictional force
proportional to the particles' velocity added on top of the usual interaction. We find this
procedure more effective than a direct energy minimization, probably owing to the floppiness of the
network translating into large nearly flat regions in the potential landscape. Denoting with
$\mathbb{M}$ the relaxation procedure, we can formally write $\tilde{\mathscr{C}}_0 = \mathbb{M}
\mathscr{C}_0$. 

We then apply a small, instantaneous, homogeneous shear deformation $\mathbb{T}_\varepsilon$ to the
simulation box, obtaining the configuration $\mathscr{C}_1 =
\mathbb{T}_\varepsilon\tilde{\mathscr{C}}_0$. Finally, we relax the configuration to the nearest inherent
structure, obtaining $\tilde{\mathscr{C}}_1 = \mathbb{M} \mathscr{C}_1$. The non-affine displacement
field induced by the shear strain $\varepsilon$ is defined as $\{\mathbf{d}_i\}_\varepsilon =
\tilde{\mathscr{C}}_1 - \mathscr{C}_1 = \mathbb{M}\mathbb{T}_\varepsilon \tilde{\mathscr{C}}_0 -
\mathbb{T}_\varepsilon \tilde{\mathscr{C}}_0$, where the difference between configurations is to be
understood as the set of differences of the positions of corresponding particles. This basic step
can be iterated any number of times to obtain the non-affine displacement field corresponding to a
strain $n\varepsilon$: $\{\mathbf{d}_i\}_{n\varepsilon} = (\mathbb{M}\mathbb{T}_\varepsilon)^n
\tilde{\mathscr{C}}_0 - \mathbb{T}_\varepsilon^n \tilde{\mathscr{C}}_0$.

We have performed the analysis on six independent gel configurations with $N=4000$ particles,
deforming with steps of $\varepsilon=1\%$ strain, up to a final strain $n\varepsilon=15\%$. The
configurations are the same as the ones used for the iso-configurational analysis. We quantify the
correlation between the magnitude of the non-affine displacements and the local density of
crosslinks by means of Pearson's coefficient of correlation
$r(\{\Vert\mathbf{d}_i\Vert\}_{n\varepsilon}; \ci)$. The correlation coefficients corresponding
to different strains are plotted in Fig.~\ref{fig:nad_corr}. We find a consistent negative
correlation, meaning that the larger non-affine displacements happen preferentially in regions with
low density of crosslinks.

%%\end{document}

%\bibliography{prl_colombo}
%\bibliographystyle{nmag}


\end{document}